\documentclass[reprint,prd,a4paper,showpacs,amsmath,superscriptaddress]{revtex4-1}

\usepackage{graphicx}
\usepackage[]{natbib}
\usepackage{dcolumn}

\begin{document}

\title{First all-sky search for continuous gravitational waves from unknown sources in binary systems}
\author{%
J.~Aasi$^{1}$,
B.~P.~Abbott$^{1}$,
R.~Abbott$^{1}$,
T.~Abbott$^{2}$,
M.~R.~Abernathy$^{1}$,
T.~Accadia$^{3}$,
F.~Acernese$^{4,5}$,
K.~Ackley$^{6}$,
C.~Adams$^{7}$,
T.~Adams$^{8}$
% ...
}\noaffiliation

\affiliation {LIGO - California Institute of Technology, Pasadena, CA 91125, USA }
\affiliation {Louisiana State University, Baton Rouge, LA 70803, USA }
\affiliation {Laboratoire d'Annecy-le-Vieux de Physique des Particules (LAPP), Universit\'e de Savoie, CNRS/IN2P3, F-74941 Annecy-le-Vieux, France }
\affiliation {INFN, Sezione di Napoli, Complesso Universitario di Monte S.Angelo, I-80126 Napoli, Italy }
\affiliation {Universit\`a di Salerno, Fisciano, I-84084 Salerno, Italy }
\affiliation {University of Florida, Gainesville, FL 32611, USA }
\affiliation {LIGO - Livingston Observatory, Livingston, LA 70754, USA }
\affiliation {Cardiff University, Cardiff, CF24 3AA, United Kingdom }
\affiliation {Albert-Einstein-Institut, Max-Planck-Institut f\"ur Gravitationsphysik, D-30167 Hannover, Germany }
\email{evan.goetz@aei.mpg.de}
\affiliation {Nikhef, Science Park, 1098 XG Amsterdam, The Netherlands }
\affiliation {LIGO - Massachusetts Institute of Technology, Cambridge, MA 02139, USA }
\affiliation {Instituto Nacional de Pesquisas Espaciais, 12227-010 - S\~{a}o Jos\'{e} dos Campos, SP, Brazil }
\affiliation {Inter-University Centre for Astronomy and Astrophysics, Pune - 411007, India }
\affiliation {Tata Institute for Fundamental Research, Mumbai 400005, India }
\affiliation {Syracuse University, Syracuse, NY 13244, USA }
\affiliation {University of Wisconsin--Milwaukee, Milwaukee, WI 53201, USA }
\affiliation {Leibniz Universit\"at Hannover, D-30167 Hannover, Germany }
\affiliation {INFN, Sezione di Pisa, I-56127 Pisa, Italy }
\affiliation {Universit\`a di Siena, I-53100 Siena, Italy }
\affiliation {Stanford University, Stanford, CA 94305, USA }
\affiliation {The University of Mississippi, University, MS 38677, USA }
\affiliation {California State University Fullerton, Fullerton, CA 92831, USA }
\affiliation {INFN, Sezione di Roma, I-00185 Roma, Italy }
\affiliation {University of Birmingham, Birmingham, B15 2TT, United Kingdom }
\affiliation {Albert-Einstein-Institut, Max-Planck-Institut f\"ur Gravitationsphysik, D-14476 Golm, Germany }
\affiliation {Montana State University, Bozeman, MT 59717, USA }
\affiliation {European Gravitational Observatory (EGO), I-56021 Cascina, Pisa, Italy }
\affiliation {LIGO - Hanford Observatory, Richland, WA 99352, USA }
\affiliation {SUPA, University of Glasgow, Glasgow, G12 8QQ, United Kingdom }
\affiliation {APC, AstroParticule et Cosmologie, Universit\'e Paris Diderot, CNRS/IN2P3, CEA/Irfu, Observatoire de Paris, Sorbonne Paris Cit\'e, 10, rue Alice Domon et L\'eonie Duquet, F-75205 Paris Cedex 13, France }
\affiliation {Columbia University, New York, NY 10027, USA }
\affiliation {Universit\`a di Pisa, I-56127 Pisa, Italy }
\affiliation {CAMK-PAN, 00-716 Warsaw, Poland }
\affiliation {Astronomical Observatory Warsaw University, 00-478 Warsaw, Poland }
\affiliation {INFN, Sezione di Genova, I-16146 Genova, Italy }
\affiliation {Universit\`a degli Studi di Genova, I-16146 Genova, Italy }
\affiliation {San Jose State University, San Jose, CA 95192, USA }
\affiliation {Faculty of Physics, Lomonosov Moscow State University, Moscow 119991, Russia }
\affiliation {LAL, Universit\'e Paris-Sud, IN2P3/CNRS, F-91898 Orsay, France }
\affiliation {NASA/Goddard Space Flight Center, Greenbelt, MD 20771, USA }
\affiliation {University of Western Australia, Crawley, WA 6009, Australia }
\affiliation {Department of Astrophysics/IMAPP, Radboud University Nijmegen, P.O. Box 9010, 6500 GL Nijmegen, The Netherlands }
\affiliation {Universit\'e Nice-Sophia-Antipolis, CNRS, Observatoire de la C\^ote d'Azur, F-06304 Nice, France }
\affiliation {Institut de Physique de Rennes, CNRS, Universit\'e de Rennes 1, F-35042 Rennes, France }
\affiliation {Laboratoire des Mat\'eriaux Avanc\'es (LMA), IN2P3/CNRS, Universit\'e de Lyon, F-69622 Villeurbanne, Lyon, France }
\affiliation {Washington State University, Pullman, WA 99164, USA }
\affiliation {INFN, Sezione di Perugia, I-06123 Perugia, Italy }
\affiliation {INFN, Sezione di Firenze, I-50019 Sesto Fiorentino, Firenze, Italy }
\affiliation {Universit\`a degli Studi di Urbino 'Carlo Bo', I-61029 Urbino, Italy }
\affiliation {University of Oregon, Eugene, OR 97403, USA }
\affiliation {Laboratoire Kastler Brossel, ENS, CNRS, UPMC, Universit\'e Pierre et Marie Curie, F-75005 Paris, France }
\affiliation {VU University Amsterdam, 1081 HV Amsterdam, The Netherlands }
\affiliation {University of Maryland, College Park, MD 20742, USA }
\affiliation {University of Massachusetts - Amherst, Amherst, MA 01003, USA }
\affiliation {Universitat de les Illes Balears, E-07122 Palma de Mallorca, Spain }
\affiliation {Universit\`a di Napoli 'Federico II', Complesso Universitario di Monte S.Angelo, I-80126 Napoli, Italy }
\affiliation {Canadian Institute for Theoretical Astrophysics, University of Toronto, Toronto, Ontario, M5S 3H8, Canada }
\affiliation {Tsinghua University, Beijing 100084, China }
\affiliation {University of Michigan, Ann Arbor, MI 48109, USA }
\affiliation {Rochester Institute of Technology, Rochester, NY 14623, USA }
\affiliation {INFN, Sezione di Roma Tor Vergata, I-00133 Roma, Italy }
\affiliation {National Tsing Hua University, Hsinchu Taiwan 300 }
\affiliation {Charles Sturt University, Wagga Wagga, NSW 2678, Australia }
\affiliation {Caltech-CaRT, Pasadena, CA 91125, USA }
\affiliation {Pusan National University, Busan 609-735, Korea }
\affiliation {Australian National University, Canberra, ACT 0200, Australia }
\affiliation {Carleton College, Northfield, MN 55057, USA }
\affiliation {INFN, Gran Sasso Science Institute, I-67100 L'Aquila, Italy }
\affiliation {Universit\`a di Roma Tor Vergata, I-00133 Roma, Italy }
\affiliation {Universit\`a di Roma 'La Sapienza', I-00185 Roma, Italy }
\affiliation {University of Brussels, Brussels 1050 Belgium }
\affiliation {Sonoma State University, Rohnert Park, CA 94928, USA }
\affiliation {Embry-Riddle Aeronautical University, Prescott, AZ 86301, USA }
\affiliation {The George Washington University, Washington, DC 20052, USA }
\affiliation {University of Cambridge, Cambridge, CB2 1TN, United Kingdom }
\affiliation {Northwestern University, Evanston, IL 60208, USA }
\affiliation {The University of Texas at Brownsville, Brownsville, TX 78520, USA }
\affiliation {University of Minnesota, Minneapolis, MN 55455, USA }
\affiliation {The University of Sheffield, Sheffield S10 2TN, United Kingdom }
\affiliation {Wigner RCP, RMKI, H-1121 Budapest, Konkoly Thege Mikl\'os\'ut 29-33, Hungary }
\affiliation {University of Sannio at Benevento, I-82100 Benevento, Italy }
\affiliation {INFN, Gruppo Collegato di Trento, I-38050 Povo, Trento, Italy }
\affiliation {Universit\`a di Trento, I-38050 Povo, Trento, Italy }
\affiliation {Montclair State University, Montclair, NJ 07043, USA }
\affiliation {The Pennsylvania State University, University Park, PA 16802, USA }
\affiliation {MTA E\"otv\"os University, `Lendulet' A. R. G., Budapest 1117, Hungary }
\affiliation {Universit\`a di Perugia, I-06123 Perugia, Italy }
\affiliation {Rutherford Appleton Laboratory, HSIC, Chilton, Didcot, Oxon, OX11 0QX, United Kingdom }
\affiliation {Perimeter Institute for Theoretical Physics, Ontario, N2L 2Y5, Canada }
\affiliation {American University, Washington, DC 20016, USA }
\affiliation {University of Adelaide, Adelaide, SA 5005, Australia }
\affiliation {Raman Research Institute, Bangalore, Karnataka 560080, India }
\affiliation {Korea Institute of Science and Technology Information, Daejeon 305-806, Korea }
\affiliation {Bia{\l }ystok University, 15-424 Bia{\l }ystok, Poland }
\affiliation {University of Southampton, Southampton, SO17 1BJ, United Kingdom }
\affiliation {IISER-TVM, CET Campus, Trivandrum Kerala 695016, India }
\affiliation {Institute of Applied Physics, Nizhny Novgorod, 603950, Russia }
\affiliation {Seoul National University, Seoul 151-742, Korea }
\affiliation {Hanyang University, Seoul 133-791, Korea }
\affiliation {IM-PAN, 00-956 Warsaw, Poland }
\affiliation {NCBJ, 05-400\'Swierk-Otwock, Poland }
\affiliation {Institute for Plasma Research, Bhat, Gandhinagar 382428, India }
\affiliation {The University of Melbourne, Parkville, VIC 3010, Australia }
\affiliation {INFN, Sezione di Padova, I-35131 Padova, Italy }
\affiliation {Monash University, Victoria 3800, Australia }
\affiliation {SUPA, University of Strathclyde, Glasgow, G1 1XQ, United Kingdom }
\affiliation {ESPCI, CNRS, F-75005 Paris, France }
\affiliation {Argentinian Gravitational Wave Group, Cordoba Cordoba 5000, Argentina }
\affiliation {Universit\`a di Camerino, Dipartimento di Fisica, I-62032 Camerino, Italy }
\affiliation {The University of Texas at Austin, Austin, TX 78712, USA }
\affiliation {Southern University and A\&M College, Baton Rouge, LA 70813, USA }
\affiliation {College of William and Mary, Williamsburg, VA 23187, USA }
\affiliation {IISER-Kolkata, Mohanpur, West Bengal 741252, India }
\affiliation {National Institute for Mathematical Sciences, Daejeon 305-390, Korea }
\affiliation {Hobart and William Smith Colleges, Geneva, NY 14456, USA }
\affiliation {RRCAT, Indore MP 452013, India }
\affiliation {SUPA, University of the West of Scotland, Paisley, PA1 2BE, United Kingdom }
\affiliation {Institute of Astronomy, 65-265 Zielona G\'ora, Poland }
\affiliation {Indian Institute of Technology, Gandhinagar Ahmedabad Gujarat 382424, India }
\affiliation {Instituto de F\'\i sica Te\'orica, University Estadual Paulista/International Center for Theoretical Physics-South American Institue for Research, S\~ao Paulo SP 01140-070, Brazil }
\affiliation {Andrews University, Berrien Springs, MI 49104, USA }
\affiliation {Trinity University, San Antonio, TX 78212, USA }
\affiliation {University of Washington, Seattle, WA 98195, USA }
\affiliation {Southeastern Louisiana University, Hammond, LA 70402, USA }
\affiliation {Abilene Christian University, Abilene, TX 79699, USA }
\collaboration{LIGO Scientific Collaboration and Virgo Collaboration}
\noaffiliation

\begin{abstract}
We present the first results of an all-sky search for continuous gravitational waves from unknown spinning neutron stars in binary systems using LIGO and Virgo data. Using a specially developed analysis program, the TwoSpect algorithm, the search was carried out on data from the sixth LIGO Science Run and the second and third Virgo Science Runs. The search covers a range of frequencies from 20~Hz to 520~Hz, a range of orbital periods from 2 to $\sim$2,254~h and a frequency- and period-dependent range of frequency modulation depths from 0.277 to 100~mHz. This corresponds to a range of projected semi-major axes of the orbit from $\sim$$0.6\times10^{-3}$~ls to $\sim$6,500~ls assuming the orbit of the binary is circular. While no plausible candidate gravitational wave events survive the pipeline, upper limits are set on the analyzed data. The most sensitive 95\% confidence upper limit obtained on gravitational wave strain is $2.3\times10^{-24}$ at 217~Hz, assuming the source waves are circularly polarized. Although this search has been optimized for circular binary orbits, the upper limits obtained remain valid for orbital eccentricities as large as 0.9. In addition, upper limits are placed on continuous gravitational wave emission from the low-mass x-ray binary Scorpius X-1 between 20~Hz and 57.25~Hz.
\end{abstract}


\maketitle

\section{Introduction}
Rapidly rotating, non-axisymmetric neutron stars are predicted to emit continuous, nearly monochromatic gravitational waves. Using data from previous LIGO (Laser Interferometer Gravitational wave Observatory)~\citep{LIGOifo2009} and Virgo~\citep{VirgoDetector} science runs, other all-sky searches have been carried out for continuous gravitational wave signals from isolated, spinning neutron stars. Past all-sky searches include two different searches on LIGO Science Run 2 (S2) data~\citep{S2cwSearch,S2Hough}; three StackSlide-like search algorithms~\citep{StackSlideMethod,S4PSH} and the first Einstein@Home distributed computing search~\citep{S4EatH} on LIGO Science Run 4 (S4) data; and a PowerFlux search~\citep{EarlyS5PowerFlux,FullS5PowerFlux} and Einstein@Home search~\citep{EarlyS5EatH,FullS5EatH} carried out on LIGO Science Run 5 (S5) data. 
None of these searches directly addressed continuous waves from a neutron star in an unknown binary system and none had appreciable sensitivity to such sources because of orbital modulation effects discussed below. Previous searches have been carried out, however, for a signal from the known low-mass x-ray binary system, Scorpius X-1, where the binary orbital parameters are reasonably constrained. One method used LIGO S2 data~\citep{S2cwSearch}, and a different method used LIGO S4 data~\citep{S4directionalStochastic} and S5 data~\citep{S5directionalStochastic}. This article presents an explicit search for continuous waves from unknown neutron stars in binary systems, as well as a directed search for gravitational waves from Scorpius X-1.

Although the waves emitted by a spinning neutron star are nearly monochromatic, a gravitational wave detector located on Earth would observe a frequency-modulated signal caused by the motion of Earth~\citep{JKS1998}. Additionally, if such a source is located in a binary system, then the observed waves will have a frequency modulation imposed by the motion of the source in the binary system~\citep{VeccBinary,TwoSpectMethod}. Together, these frequency modulations make searches for unknown, spinning binary neutron stars emitting continuous gravitational waves computationally demanding.

Previous searches, not accounting for the orbital modulations, would have been much less sensitive to stars in binary systems with induced frequency modulation amplitudes much greater than the frequency spacing between search templates ($\sim$$5-500$~$\mu$Hz, depending on the search method). In addition, while a very large amplitude continuous gravitational wave source in a binary system could produce outliers in other searches, the follow up of those outliers would have likely rejected them because they do not follow the expected frequency evolution of an isolated source of continuous gravitational waves. Regardless, the upper limits set by these searches would be unreliable for sources in binary systems with significant frequency modulation.

Explicitly including the effects due to unknown binary orbital parameters in the other all-sky StackSlide-like~\citep{S2Hough,S4PSH,EarlyS5PowerFlux,FullS5PowerFlux} or Einstein@Home~\citep{S4EatH,EarlyS5EatH,FullS5EatH} algorithms would be computationally prohibitive. Thus, new methods are required to perform such a search with limited computational resources~\citep{TwoSpectMethod,PolynomialSearch}. These new techniques require some sacrifice of strain sensitivity to gravitational waves in order to significantly reduce the computational demands of such a search.

One such algorithm, called TwoSpect~\citep{TwoSpectMethod}, has been developed, implemented, and a search carried out with it using recently collected LIGO and Virgo data. The TwoSpect algorithm relies on the periodic nature of the frequency modulation caused by the binary orbit. Spectrograms of gravitational wave detector data are created after correcting for the Earth's known rotation and orbital motion, and then Fourier transformations of each frequency bin of the barycentered spectrogram are computed. These successive Fourier transforms enable efficient detection of frequency modulated signals because the modulation has fixed periodicity. Although optimized for circular orbits, the methodology used to obtain upper limits on source strengths remains sensitive for eccentricities as large as 0.9.

This article is organized as follows: section~\ref{sec:astroSources} discusses neutron stars in binary systems and the assumed signal model; section~\ref{sec:detectors} briefly describes the LIGO and Virgo gravitational wave detectors; section~\ref{sec:method} discusses the TwoSpect method; section~\ref{sec:analysis} describes the analysis of the detector data, and section~\ref{sec:results} gives the results of the analysis; section~\ref{sec:conclusions} summarizes the conclusions of this work.

\section{\label{sec:astroSources}Astrophysical sources}
Spinning neutron stars in binary systems are attractive sources in searches for continuous gravitational waves because accretion from a companion may cause an asymmetrical quadrupole moment of inertia of the spinning neutron star. Many mechanisms have been proposed where gravitational wave emission continues after accretion of material has subsided. For instance, the magnetic field of the neutron star can guide the accretion flow to `hot spots' which could build up the neutron star ellipticity close to that allowed by the maximum breaking strain of the crust~\citep{Bildsten98}, with possibly sustained localized mass accumulation~\citep{VigeliusMelatos2009a}, depending on nuclear equation of state~\citep{Priymak2011}, material sinking~\citep{Wette2010}, resistive relaxation~\citep{VigeliusMelatos2009b} and magnetic bottling stability~\citep{VigeliusMelatos2008}. In addition, magnetic fields could create non-axisymmetric deformations of the neutron star interior~\citep{MagneticMountain}, or r-mode oscillations of the neutron star might be sustained causing the star to emit gravitational waves~\citep{MSPsRmodes,AccretingAndYoungNSrModes}.

Accreting neutron stars can be spun up by acquiring angular momentum from the infalling matter. All-sky surveys of millisecond pulsars have found that no neutron stars are spinning close to their predicted break-up frequency ($\nu\sim1400$ Hz)~\citep{atnfcatalog}. Since the observed spin frequency range of actively accreting millisecond pulsars is $180\,\text{Hz}<\nu<600\,\text{Hz}$~\cite{AccretingMSPreview}, there may be a competing mechanism preventing the spin-up of the neutron star from reaching the break-up frequency.

It has been postulated that there exists a torque balance between the accretion spin-up and the gravitational emission spin-down~\citep{PandPtorqueBalance,Wagoner84,Bildsten98}. In such a case, those neutron stars accreting at the highest rates should have the highest gravitational wave emissions. Using this relation to balance spin-down of gravitational wave emission with x-ray luminosity (a measure of the accretion rate), the dimensionless gravitational wave amplitude, $h_0$, is given by
\begin{eqnarray}
 h_0 &=& 2.7\times10^{-26} \left(\frac{f}{800\,\text{Hz}}\right)^{-1/2} \times \nonumber \\*
 &&\left(\frac{F_\text{x}}{3.9\times10^{-7}\,\text{erg}\,\text{cm}^{-2}\,\text{s}^{-1}}\right)^{1/2}\,,
\end{eqnarray}
where $f$ is the gravitational wave frequency and $F_\text{x}$ is the average bolometric x-ray flux detected at the Earth. The x-ray luminosity is scaled to the average bolometric flux of Scorpius X-1 (Sco X-1). If r-mode instabilities are driven by the accretion of material, then the gravitational wave amplitude could be increased as~\citep{Owen10}
\begin{eqnarray}
 h_0 &=& 3.3\times10^{-26} \left(\frac{f}{800\,\text{Hz}}\right)^{-1/2} \times \nonumber \\*
 &&\left(\frac{F_x}{3.9\times10^{-7}\,\text{erg}\,\text{cm}^{-2}\,\text{s}^{-1}}\right)^{1/2}\,.
\end{eqnarray}

\subsection{Gravitational wave signal model}
The expected waveform of a non-axisymmetric spinning neutron star observed by a gravitational wave interferometer is
\begin{eqnarray}
h(t)&  =& h_0F_+(t,\alpha,\delta,\psi)\frac{1+\cos^2(\iota)}{2}\cos\left[\Phi(t)\right] + \nonumber \\*
& & h_0F_\times(t,\alpha,\delta,\psi)\cos(\iota)\sin\left[\Phi(t)\right]\,,
\end{eqnarray}
where $F_+$ and $F_\times$ are the detector response functions (antenna patterns) to `plus' and `cross' polarized gravitational waves, $\alpha$ and $\delta$ are the right ascension and declination of a particular sky location, $\psi$ is the polarization angle of the waves, $\iota$ is the inclination angle of the neutron star rotational axis to the line of sight, and $\Phi(t)$ is the phase evolution of the gravitational wave signal. The assumed instantaneous phase evolution is given by
\begin{eqnarray}
\Phi(t) & =& \Phi_0 + 2\pi f_0(t-t_\text{ref}) + \nonumber \\*
 & & 2\pi\Delta f_\text{obs}\sin[\Omega (t-t_\text{asc})]/\Omega \,,
\label{eq:phaseEvolution}
\end{eqnarray}
where $t$ is the time in the Solar System barycenter (SSB) frame, $\Phi_0$ and $f_0$ are phase and frequency, respectively, determined at reference time $t_\text{ref}$, and $t_\text{asc}$ is a given time of the orbital ascending node. The observed frequency modulation depth $\Delta f_\text{obs}$ and period of frequency modulation $P=2\pi\Omega^{-1}$ are caused by the motion of the source.

We assume that any spindown effects---$2\pi\dot{f}(t-t_0)^2$ and higher order terms---in the phase evolution of the source are negligible during the observation time and that the orbit is circular and non-relativistic. Electromagnetic observational evidence has shown that pulsars in binary systems typically have very small spindowns, $|\dot{\nu}|<10^{-15}$~Hz~s$^{-1}$ (although larger spindown could imply larger-amplitude gravitational wave emission), and also have nearly circular orbits. It may be possible, however, that a neutron star in a binary system with a small spindown value could be a strong emitter of gravitational radiation (for example, the neutron star is in torque balance equilibrium). Even though a circular orbit phase model has been assumed, the detection algorithm is sensitive to the more general case of an eccentric orbit.

The gravitational wave amplitude for a non-axisymmetric spinning neutron star with $l=m=2$ mass quadrupole moment is
\begin{equation}
h_0 = \frac{16\pi^2G}{c^4}\frac{I\epsilon\nu^2}{d}\,,
\end{equation}
where $G$ is the gravitational constant, $c$ is the speed of light in a vacuum, $I$ is the principal moment of inertia with respect to the spin axis, $\epsilon$ is the equatorial ellipticity of the neutron star, $\nu$ is the rotational frequency of the neutron star, and $d$ is the distance to the neutron star. A spinning neutron star will emit continuous gravitational waves with frequency $f_0=2\nu$.

The observed modulation depth is related to the maximum modulation depth, $\Delta f_\text{max}$, by
\begin{equation}
\Delta f_\text{obs} = \Delta f_\text{max}\sin i\,,
\end{equation}
where $i$ is the inclination angle of the binary orbital plane with respect to the vector that points from the detector to the sky position. Assuming a circular, non-relativistic orbit, the maximum observable Doppler shift will occur for an edge-on observed system with the modulation depth given by~\citep{TwoSpectMethod}
\begin{eqnarray}
\Delta f_\text{max} &\simeq& 1.82 \, \left(\frac{f_0}{1\,\text{kHz}}\right) \left(\frac{M_\text{NS}}{1.4\,M_{\odot}}\right)^{1/3} \times \nonumber \\*
& & \left(\frac{P}{2\,\text{h}}\right)^{-1/3} \left[\frac{q}{(1+q)^{2/3}}\right]\,\text{Hz}\,,
\end{eqnarray}
where $M_\text{NS}$ is the mass of the neutron star and $q\equiv M_2/M_\text{NS}$ is the mass ratio of the companion mass to the neutron star mass.

Alternatively, the observed modulation depth for a circular, non-relativistic orbit can be written with directly observable parameters
\begin{equation}
\Delta f_\text{obs} \simeq 0.8727 \left(\frac{f_0}{1\,\text{kHz}}\right) \left(\frac{P}{2\,\text{h}}\right)^{-1} \left(\frac{a\sin i}{1\,\text{ls}}\right) \,\text{Hz}\,,
\label{eq:moddepth2}
\end{equation}
where $a\sin i$ is the projected semi-major axis (the projected radius of the orbit since we are concerned with nearly circular orbits) in units of light seconds (ls). Given a wide range of realistic orbital parameters, Eq.~(\ref{eq:moddepth2}) shows that one must search frequency modulation depths easily reaching 1~Hz or greater, to cover the full range of possible binary systems.

\section{\label{sec:detectors}LIGO and Virgo detectors}
Data taken in 2009-2010 with the 4-km-long `enhanced' LIGO detectors~\citep{LIGOifo2009} and the 3-km-long Virgo detector~\citep{VirgoDetector} were used in this analysis. The LIGO and Virgo detectors are both power-recycled Michelson interferometers with Fabry-Perot arm cavities.

Following the fifth LIGO Science Run (S5), a number of upgrades were made to the `initial' LIGO 4-km-long interferometers (H1 in Hanford, Washington, and L1 in Livingston Parish, Louisiana). Most substantially: 1) the initial 10~W laser was upgraded to a new 35~W laser, 2) an `output mode cleaner' was installed at the output port of the interferometer, 3) the radio-frequency detection scheme (heterodyne) was changed to a DC detection scheme (homodyne), and 4) the detection opto-electronics were moved to an in-vacuum, actively-stabilized optical table to reduce seismic motion affecting the read-out optics and electronics. These upgrades constituted the enhanced LIGO interferometers~\citep{eLIGO}.

Following the first Virgo Science Run (VSR1), several upgrades were made to improve the sensitivity of the detector for the subsequent second and third Science Runs. The main enhancements to the detector included: 1) upgrading to a new 25~W laser, 2) installation of a thermal compensation system to reduce thermal effects of laser power absorption in the main interferometer mirrors, 3) replacement of read-out and control electronics with lower-noise components, and 4) between the second and third Science Runs, new, monolithic, low-loss, fused silica suspensions were installed on the main interferometer mirrors~\citep{eVirgo}.

During the period of 7 July 2009 to 20 October 2010, the two enhanced LIGO 4-km interferometers, H1 and L1, had their sixth Science Run (S6), while the Virgo interferometer had its second Science Run (VSR2) concurrently from 7 July 2009 to 8 January 2010 and third Science Run (VSR3) from 11 August 2010 to 19 October 2010. The increased input laser power of the upgraded LIGO detectors decreased the noise above 200~Hz compared to S5 by a factor of $\sim$2, with more modest improvements below 200~Hz. The Virgo detector has a better sensitivity compared to the enhanced LIGO detectors below $\sim$50~Hz, but worse sensitivity at higher frequencies.

\section{\label{sec:method}TwoSpect algorithm}
The details of the TwoSpect method have been described previously~\citep{TwoSpectMethod}. We briefly summarize the algorithm here. Short segments (30~min or less) of gravitational wave detector data are Fourier transformed (so-called Short Fourier Transforms, or SFTs) using the FFTW (Fastest Fourier Transform in the West) algorithm~\citep{FFTW}, and the power of each Fourier coefficient is computed. Next, each SFT is weighted according to the noise present in the SFT, and by the antenna pattern of the detector (the sensitivity) to a given sky location at the particular time that the SFT data were recorded. Time spectrograms of SFTs over a narrow frequency band ($\sim$1~Hz) are created such that the frequency shift caused by Earth's motion is removed by sliding each SFT by an appropriate amount for a specific sky location. Then, for each such spectrogram, the Fourier transform of each frequency bin's powers as a function of time is computed and, from these Fourier coefficients, the power spectra of the second Fourier transform is determined.

The TwoSpect search for gravitational waves is hierarchically organized into two stages. First, a non-template-based algorithm searches the doubly-Fourier-transformed data for interesting regions of parameter space that exceed a specific threshold value. Second, the interesting regions of parameter space are subjected to template-based tests in order to confirm or reject specific outliers. Whether or not an outlier has been found, an upper limit on gravitational wave amplitude is placed at each sky location.

\subsection{\label{sec:dataPrep}Data preparation}
The S6 and VSR2/3 data sets, each defined here with a length $T_\text{obs}=40551300$~s, are divided into segments of length $T_\text{SFT}=1800$~s. Each sequential segment overlaps the preceding segment by 50\%, and each of these segments of data is windowed using the Hann window function, to suppress signal leakage into other frequency bins, before the Fourier transform is computed. The windowed Fourier transform is defined as
\begin{equation}
\tilde{s}_k = \frac{\Delta t}{C} \sum_{j=0}^{M-1}w_js_je^{-2\pi \text{i}jk/M}\,,
\end{equation}
where $k=0,1,2,\ldots,(M-1)$, $\Delta t=T_\text{SFT}/M$ is the sampling interval, the window function is $w_j = 0.5[1-\cos(2\pi j/M)]$, and $C=(\sum_{j=0}^{M-1}w_j^2/M)^{1/2}=(3/8)^{1/2}$. Physical frequency $f_k=k/T_\text{SFT}$ corresponds to $0\leq k\leq M/2$. The `power' in bin $k$ of SFT $n$ is taken to be
\begin{equation}
P_k^n = \frac{2\left|\tilde{s}_k^n\right|^2}{T_\text{SFT}}\,.
\end{equation}

The SFTs are adjusted for the changing detector velocity with respect to a fixed sky location by shifting SFT bins to correct for this effect in the same manner as other StackSlide-like algorithms~\citep{StackSlideMethod,S4PSH}. A sequence of $n$ (shifted) SFT powers are weighted and normalized by
\begin{equation}
 \widetilde{P}_k^n = \frac{F_n^2(P_k^n - \langle P_k\rangle^n)}{(\langle P_k\rangle^n)^2}\left[\sum_{n^\prime}^N\frac{F_{n^\prime}^4}{(\langle P_k\rangle^{n^\prime})^2}\right]^{-1}\,,
\label{eq:Pmaster}
\end{equation}
where angle brackets $\langle\rangle$ indicate the running mean value over the inner index---the frequency bins, $k$, to estimate the noise background---and, assuming a circularly-polarized gravitational wave,
\begin{equation}
 F^2(t,\alpha,\delta)=F_{+}^2(t,\alpha,\delta)+F_{\times}^2(t,\alpha,\delta)\,.
\end{equation}
The dependence on $\psi$ has been omitted because $F^2$ has no $\psi$ dependence for circular polarization. Hence, particular SFTs that have low noise or for which the detector is favorably oriented to a sky position are weighted more heavily than SFTs that have high noise or for which the detector is unfavorably oriented.

The running mean values of the noise background are calculated from the running median values~\citep{MedianEstimate} of the SFT powers. The running median is converted to a mean value (assuming the $P_k^n$ values follow an exponential distribution) including a bias factor for this analysis of a running median of 101 bins~\citep{S4PSH}. The running mean values are an estimate of the smoothly-varying detector noise background that avoids biases from sharp spectral features of the detector noise (lines) and potential signals.

The Fourier transform of Eq.~(\ref{eq:Pmaster}) is then computed for each frequency bin $k$, and normalized such that the expectation value of the second Fourier transform in the presence of noise is equal to 1. For frequency bin $k$, the power as a function of second Fourier transform frequency, $f^\prime$, is written as
\begin{equation}
Z_k(f^\prime) = \frac{\left|\mathcal{F}\left[\widetilde{P}_k^n\right]\right|^2}{\langle \lambda(f^\prime)\rangle}\,,
\label{eq:2ndFFT}
\end{equation}
where $\mathcal{F}$ denotes a Fourier transform, and $\langle\lambda(f^\prime)\rangle$ is the mean of the background noise estimate of the second Fourier transform. The values of $\lambda(f^\prime)$ are determined by Monte Carlo simulation using the noise estimates established from the SFTs and assuming the noise in the SFTs is due to Gaussian noise alone. The distribution of $Z_k(f^\prime)$ values from a Gaussian-noise time-series follows a $\chi^2$ distribution with two degrees of freedom and mean of 1.0 to a good approximation, as discussed in~\citep{TwoSpectMethod}.

Note that in this analysis, $Z_k(f^\prime)$ is directly proportional to $h^4$ because the power spectrum of SFT powers (directly proportional to $h^2$) has been computed. This means that detection statistics computed from $Z_k$ values will be directly proportional to $h^4$.

\subsection{First-stage detection statistic}
The all-sky search begins with an untemplated search algorithm, incoherent harmonic summing (IHS)~\citep{TwoSpectMethod}, to identify regions of parameter space to be searched later using templates and to set upper limits. It is useful to define a quantity that measures power at multiple harmonics of a fundamental frequency, $f^\prime$. For example, one can fold each $Z_k(f^\prime)$ an integer $j = 1\ldots S$ times to define for a single frequency bin, $k$, the following statistic:
\begin{equation}
\mathcal{V}_k = \textrm{max}\left\{\sum_{j=1}^S\left[Z_k(jf^\prime)-\lambda(jf^\prime)\right]\right\}\,.
\end{equation}
If a periodic signal is present, then the IHS algorithm will accumulate signal power from the higher harmonic frequencies into the lower harmonic frequencies. The signal-to-noise ratios of the signal bins grow $\propto$$\sqrt{S}$, provided the sequence of harmonic powers have similar SNR in the original spectra. In practice, this increase in SNR is limited by the strength of the higher signal harmonics, giving the IHS technique a practical limit of $S\sim5$ in this application.

To accumulate additional signal power, folded $Z_k$ values are summed across sequential values of $k$ according to
\begin{equation}
\mathcal{W}(k_0,f^\prime,\Delta k) = \sum_{k=k_0-\Delta k}^{k_0+\Delta k}\sum_{j=1}^S\left[Z_k(jf^\prime)-\lambda(jf^\prime)\right]\,,
\label{eq:IHSstat}
\end{equation}
before determining the maximum value. Computing $\mathcal{W}(k_0,f^\prime,\Delta k)$ `compresses' the second Fourier transformed data. Then, for a chosen $\Delta k$, the maximum value of $\mathcal{W}(k_0,f^\prime)$ is determined. As described in section~\ref{sec:dataPrep}, the values of $\mathcal{W}$ are proportional to $h^4$.

At the end of the first stage, any IHS statistic passing a threshold of a predetermined false alarm probability is passed to the second, template-based stage for more stringent follow-up tests using test values of $f$ (derived from $k_0$), $P$ (derived from $f^\prime$), and $\Delta f_\text{obs}$ (derived from $\Delta k$). Whether or not any candidates are found in the first stage, a frequentist 95\% confidence upper limit is placed based on the highest statistic found in the first stage (see section~\ref{sec:ULs}).

\subsection{Second-stage detection statistic}
The second stage of the pipeline tests candidate outliers from the first stage against templates that are based on putative signal patterns and weights in the second Fourier transform. Assume that the strain power for a putative signal is distributed among $M$ pixels of the second Fourier transform for a narrow band of SFT frequencies, with the fraction of the signal power in pixel $m_i$ equal to $w(m_i)$. A useful statistic to sum pixel powers is
\begin{equation}
 R = \frac{\sum_{i=0}^{M-1} w(m_i)[Z(m_i) - \lambda(m_i)]}{\sum_{i=0}^{M-1}[w(m_i)]^2}\,,
\label{eq:Rstat}
\end{equation}
where $Z(m_i)$ is the second Fourier transform power in pixel $m_i$ (each $m_i$ is a unique value of $k$ and $f^\prime$), $\lambda(m_i)$ is the expected noise value of pixel $m_i$ of the second Fourier transform, and the weights are normalized such that
\begin{equation}
\sum_{i=0}^{M-1} w(m_i) = 1\,,
\end{equation}
where $N$ is the total number of pixels in the region of interest of the second Fourier transform. In practice, due to computational constraints, the value of $M$ in Eq.~(\ref{eq:Rstat}) is fixed to be no larger than 500. This limit is raised in follow-up studies of particularly interesting candidates.

The weights are sorted such that $w(m_0)$ contains the greatest weight and $w(m_{M-1})$ contains the smallest weight. The weights, $w(m_i)$, are determined by using a set of templates with parameters $(f,P,\Delta f)$ using the same $T_\text{SFT}$ and $T_\text{obs}$ as the search~\citep{TwoSpectMethod}. If the input time series of data is Gaussian, white noise, then the value of $R$ is a weighted $\chi^2$ variable with up to $2M$ degrees of freedom but shifted to have zero-mean. Again, the second stage statistic, $R$, is proportional to $h^4$.

For each candidate passed to the second stage, a number of different templates are tested using the `Gaussian' template approximation~\citep{TwoSpectMethod} with orbital period values up to the fifth harmonic or sub-harmonic from the originally-identified orbital period value, as well as fractional orbital period values of 2/3, 3/4, 4/5, 3/2, 4/3, and 5/4 from the originally identified orbital period value (we refer to this mis-identification as `harmonic confusion'). From the tested templates, only the most significant candidate (see section~\ref{sec:significance}) is kept to be followed up by searching a small region of $(f,P,\Delta f)$ with both `Gaussian' templates and with more exact templates. These template tests provide more stringent requirements for rejecting noise outliers.

\subsection{\label{sec:significance}Significance of outliers}
To quantify the significance of a specific value of $R_0$, given a set of $w(m_i)$, $Z(m_i)$, and an estimate of $\lambda(m_i)$, the false alarm probability $P(R\geq R_0)$ is computed. The false alarm probability is solved using the method described in~\cite{TwoSpectMethod} applying the formulas of~\citep{Davies2}. The value computed for the false alarm probability assumes the underlying noise for each pixel is $\chi^2$ distributed with 2 degrees of freedom with mean values given by $\lambda(m_i)$. The computed false alarm probability value does not take into account testing multiple points in parameter space. Section~\ref{sec:outlierprocedure} describes how the significance is used in the follow-up of analysis outliers.

\subsection{\label{sec:ULs}Determination of upper limits}
At each sky location, the algorithm sets a frequentist 95\% confidence level upper limit based on the highest calculated IHS statistic value in the searched frequency band, over the range of orbital periods and modulation depths. Upper limits are placed at this stage using the IHS because obtaining more sensitive template-based upper limits is computationally infeasible with available resources. Only promising outliers are followed up for detection using a templated search. Even in the event of a successful detection, however, IHS upper limits remain valid (see figures~\ref{fig:ulvalidation} and \ref{fig:randPolarizationUL}). In the presence of pure Gaussian noise, the IHS statistic is a $\chi^2$ variable with $2AS$ degrees of freedom, where $A$ is the number of SFT frequency bins summed, and $S$ is the number of harmonics summed in the IHS algorithm. We wish to determine the amount of signal required such that the new IHS statistic value would exceed the highest found IHS statistic value 95\% of the time.

To find the amount of signal required, we invert the non-central $\chi^2$ cumulative distribution function (CDF) so that the appropriate non-centrality parameter, $p$, is found such that only 5\% of the distribution lies below the highest outlier value. The inversion is done using Newton's method. From the calculated value of $p$ and the expected noise background, the value is converted to a value of $h_0$ such that, 95\% of the time, the calculated value of $h_0$ is larger than any potentially present continuous gravitational wave signal in the data that has parameters within the parameter space searched by TwoSpect (see, e.g., section~\ref{sec:analysis}). The conversion factor is a simple scaling factor that relates the value of $p^{1/4}$ (recall that $\mathcal{W}$ is proportional to $h^4$) to the 95\% confidence level strain amplitude upper limit, $h_0^{95\%}$. The scaling factor is determined using injections of a wide variety of waveforms covering the parameter space searched. The all-sky upper limit in a given frequency band is then determined by selecting the largest value of $h_0^{95\%}$ from the entire set of sky-coordinates searched for that frequency band.

This method of setting upper limits has been validated with simulated software injections and provides reliable results in bands that pass the data quality requirements described in section~\ref{sec:dq}.

\section{\label{sec:analysis}Analysis of the data}
\begin{figure}
\centering
\includegraphics[width=0.5\textwidth]{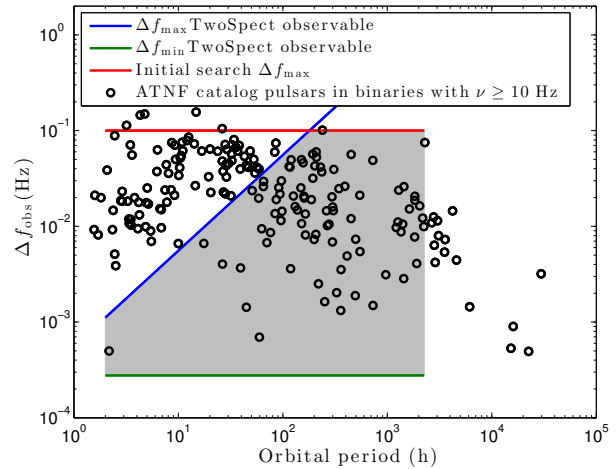}
\caption{\label{fig:paramspace}Nominal parameter space that is analyzed using the TwoSpect algorithm (shaded region). The bounding curves given by $\Delta f_\text{max}$ and $\Delta f_\text{min}$ are limitations of the analysis, while the initial search boundary of $\Delta f_\text{max}=0.1$~Hz is a choice. Data marked by circles are ATNF catalog pulsars found in binary systems with rotation frequencies $\geq$$10$~Hz (using Eq.~(\ref{eq:moddepth2}) and assuming $f_0=2\nu$).}
\end{figure}

Data from the H1 and L1 detectors' sixth Science Run (S6) and V1 detector's second and third Science Runs (VSR2 and VSR3, collectively VSR2/3) were analyzed using the TwoSpect algorithm. Each detector's data set was analyzed separately with $T_\text{obs} = 40551300$~s. An outlier from one detector is required to be coincident in parameter space with an outlier in a second detector in order to be considered a candidate signal. Figure~\ref{fig:paramspace} shows the period-modulation depth parameter space values covered in this analysis using TwoSpect.

It is assumed that the sinusoidal term in Eq.~(\ref{eq:phaseEvolution}) is constant during a single coherent observation interval, that is, the sinusoidal term is slowly evolving compared to the $f_0$ term. The signal is, therefore, assumed to be contained within one frequency bin for each coherent observation interval. This approximation restricts the orbital parameter space that can be observed: the longer a coherent observation, the more restricted the parameter space~\citep{TwoSpectMethod}. Longer coherent observation intervals, however, correspond to increased sensitivity to continuous wave signals. A trade-off is thus made in the sensitivity versus parameter space volume to be probed when conducting such a search.

LIGO S6 data from H1 and L1 were analyzed from 50~Hz to 520~Hz, covering a range of periods from 2~h to 2,254.4~h and modulation depths of 0.277~mHz to 100~mHz. Virgo VSR2/3 data were analyzed from 20~Hz to 100~Hz, over the same range of periods and modulation depths. The range of orbital periods has a lower limit determined by the coherence length of the SFTs, and an upper bound by requiring at least five orbits during the total observation time. The lower limit of modulation depths is determined by the coherence length of the SFTs, and the upper bound is chosen by covering a large region of parameter space without dramatically increasing computational costs. VSR2/3 data are only comparable to or better than LIGO S6 data in the aforementioned range of frequencies. Analyzing higher frequencies in the Virgo data would add to the total computing cost and add negligibly to the search sensitivity.

The TwoSpect program is part of the LALsuite (LIGO Analysis Library suite) software package~\citep{LALrepository}. TwoSpect is written in C and is compiled against LAL (LSC Algorithm Library), GSL (GNU Scientific Library)~\citep{gsl}, and FFTW libraries. On the LIGO computer clusters, the analysis is divided into parallel `jobs' that are run on many computers simultaneously. Each job is an instance of the TwoSpect program and analyzes a 0.25~Hz frequency band and a small sky region (typically 200 sky grid locations).

\subsection{\label{sec:dq}Data quality validation}
Ideally, the noise from a gravitational wave interferometer would be stationary Gaussian noise (in addition to any gravitational wave signal). In practice, data from the LIGO and Virgo detectors are generally stationary and nearly Gaussian on the timescale that one SFT is computed. There are occasions, however, when data must be excluded because: 1) it is known the interferometer data is corrupted (data quality flags are applied); 2) the segment of data passes data quality flags, but the data segment is non-Gaussian (a Kolmogorov-Smirnov test and/or Kuiper's test fails); or 3) sharp, stationary spectral features prevent a full analysis of the selected frequency band. Examples of sharp, stationary spectral features include: power-line harmonics (50/60~Hz), mirror suspension violin modes, and calibration lines injected into the detector by actuating one of the end mirrors. Additionally, the detectors do not operate continuously during their science runs. There are periods of downtime, or other gaps in the detector data. We describe below the techniques used to select the data to be analyzed.

\subsubsection{Science mode and data quality flags}
Periods of time when the detector was operating in the nominal `science mode' are first selected. Next, a series of quality checks of the data---known as `data quality flags'---are applied to remove times when the detector data is known to be of poor quality. Examples include when the calibration of the detector is known to be outside a tolerance range, or when there were periods of very high wind speeds (see table~\ref{tab:dataUsage})~\citep{LIGODetchar,VirgoDetchar}.
\begingroup
\squeezetable
\begin{table*}
\caption{\label{tab:dataUsage}Data usage in the S6 and VSR2/3 Science Runs.}
\begin{ruledtabular}
\begin{tabular}{l D{.}{.}{1.3}D{.}{.}{1.3}D{.}{.}{1.3}}
Duty factor condition & \multicolumn{1}{c}{H1} & \multicolumn{1}{c}{L1} & \multicolumn{1}{c}{V1} \\
\hline
Interferometer in science mode with data quality flags during the science run(s) & 0.506 & 0.463 & 0.778 \\
Interferometer in science mode with data quality flags covered by SFTs & 0.409 & 0.364 & 0.733 \\
Fraction of $T_\text{obs} = 40551300$~s covered by SFTs & 0.409 & 0.365 & 0.397 \\
Median fraction of $T_\text{obs}$ after KS and Kuiper's tests in each 0.25~Hz band & 0.383 & 0.316 & 0.366
\end{tabular}
\end{ruledtabular}
\end{table*}
\endgroup

After these checks are applied, SFTs are created. The S6 data set contains 18,435 H1 and 16,429 L1 50\% overlapping Hann-windowed SFTs with start-times occurring an integer factor of $T_{\rm SFT}/2$ from the start-time of the first SFT. The resulting duty factors are 0.409 and 0.364 for H1 and L1, respectively. The VSR2/3 data set contains 17,879 50\% overlapping Hann-windowed SFTs, corresponding to a duty factor of 0.733. SFTs consisting entirely of zeros fill in the excluded times not covered by these SFTs. The actual fraction of $T_\text{obs}$ covered by the SFTs is somewhat different. Since S6 has only a slightly longer time baseline than $T_\text{obs}$, the duty factor is nearly identical. For VSR2/3, there is a long gap in between the science runs that results in a large reduction in the fraction of $T_\text{obs}$ covered by the SFTs compared to the coverage of VSR2/3 science run time (see table~\ref{tab:dataUsage}).

\subsubsection{Kolmogorov-Smirnov test and Kuiper's test}
After the SFTs are produced, each SFT is analyzed to determine whether the distribution of the powers follows that of an expected exponential distribution. Two useful tests are the Kolmogorov-Smirnov (KS) test and Kuiper's test~\citep{NumericalRecipes}. Those SFTs which do not pass these tests are removed from the analysis, and are replaced with SFTs consisting entirely of zeros. The threshold value for each of the KS and Kuiper's test is determined from the significance level on the null hypothesis of 0.05. With this threshold, SFTs are not rejected even if they contain potential signals with the expected gravitational wave signal amplitude. For certain 0.25~Hz frequency bands for an interferometer where data coverage is less than 10\% of the total observation time due to disturbed, non-Gaussian data, no upper limits are placed in those frequency bands (see again table~\ref{tab:dataUsage}).

\subsubsection{Line detection and flagging}
Narrow spectral artifacts of terrestrial origin---also called `lines'---can potentially interfere with detections of gravitational wave signals. These disturbances are avoided by identifying potentially interfering lines (see below) and producing no further analysis of candidate signals that have interference caused by the disturbance. Upper limits are still placed, however, in frequency bands containing lines, although when the line fraction of a band exceeds 10\% of the total band, no upper limit is placed, as the noise background estimate would be untrustworthy. This problem occurs primarily in the 50 to 200~Hz region of the enhanced LIGO detectors.

Sharp spectral features are identified as an excess of power over long timescales compared to the neighboring frequency bins. The root-mean-square (RMS) power is computed for each noise-weighted SFT frequency bin as a function of time (without shifting the SFTs to account for detector motion). A running median of these RMS values is computed over the band of interest and is used to normalize the RMS values. Any normalized RMS value that exceeds an empirically-determined threshold of 1.5 is flagged as a line.

\subsubsection{Sidereal and daily modulations}
Specific orbital period frequencies corresponding to the sidereal (86164.0905~s) and daily (86400.0~s) periods and up to the third harmonic are specifically avoided in this analysis, within a tolerance of $\pm1$~second FFT frequency bin for each harmonic. These frequencies can correspond to spurious artifacts in the analysis and therefore no candidates are analyzed and no upper limits are placed at these putative binary orbital period values.

\section{\label{sec:results}Results}
The TwoSpect program produces two outputs: upper limit values and a list of outliers passing threshold tests over the parameter space searched. See section~\ref{sec:ULresults} and section \ref{sec:outlierprocedure}, respectively, for more details.

\subsection{\label{sec:ULresults}All-sky upper limit results}
\begin{figure}
\centering
\includegraphics[width=0.5\textwidth]{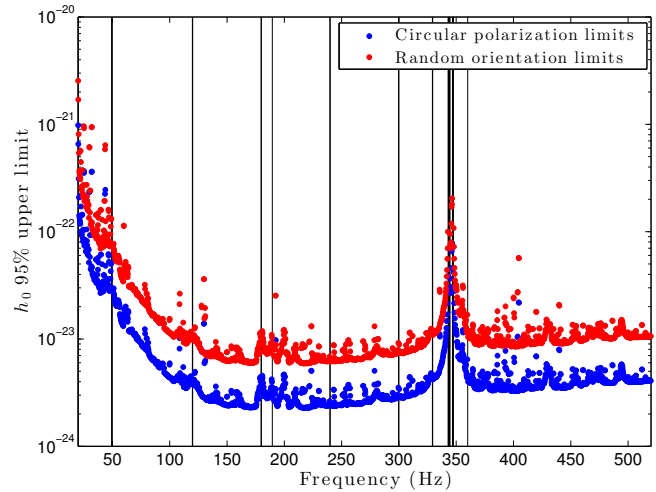}
\caption{\label{fig:ulresults}All-sky strain upper limit results of S6/VSR2-3 for continuous gravitational waves assuming the source waves are circularly polarized (blue points) or randomly polarized from randomly oriented sources (red points). The vertical black lines indicate 0.25~Hz frequency bands in which no upper limits have been placed. The smoothness of the curve is interrupted due to various instrumental artifacts, such as the violin resonances of the mirror suspensions near 350~Hz.}
\end{figure}

Upper limits are established for each interferometer separately, with a single value at each sky location. The upper limit value for a given sky location is maximized over the $(f,P,\Delta f)$ parameter space range searched. The highest upper limit value over the entire sky for a given frequency band is then selected as the overall upper limit for that frequency band in a particular interferometer. Where there is more than one detector providing an upper limit in a given frequency band, the lowest of the upper limits is taken as the overall, combined upper limit value (see figure~\ref{fig:ulresults}).

The values placed on upper limits of gravitational wave amplitude with 95\% confidence assume the best-case scenario that the un-observed gravitational waves are circularly polarized. The true astrophysical population of gravitational wave sources are expected, however, to be uniformly distributed in orientation so that the polarization of the source waves can vary over a range of values covering completely circularly-polarized waves, to completely linearly-polarized waves. In the latter case, a multiplicative scale factor of $\approx$3.3 should be applied to the results shown in figure~\ref{fig:ulresults}. In the case of random pulsar orientations, however, a scale factor of $\approx$2.6 is applied to the circular polarization results as shown in figure~\ref{fig:ulresults}. Additionally, these upper limits are valid only if the source has a spindown value $|\dot{f}|\leq1\times10^{-10}$~Hz/s.

If the source is in an eccentric orbit with its companion, then the assumed phase evolution can be a poor approximation to the true phase evolution of the source. However, whether an orbit is circular or eccentric, the modulated signal will spend more time in certain frequency bins within the modulation band (generating a stronger signal at those frequencies) and traverse other bins more rapidly (generating a weaker signal). The IHS statistic is relatively insensitive to the details of the shape of the modulation, responding only to its periodic structure in the second Fourier transform. Upper limits are still valid in this case, even for orbital eccentricities up to 0.9. An outlier caused by a gravitational wave source in an eccentric orbit may have poorly reconstructed signal parameters compared to the true source parameters.

The upper limits presented above assume that a putative signal could take any parameter values in the ranges searched. Suppose there are many  different signals contained in the data, all of them having the same $h_0$ value but with $\Delta f_\text{obs}$ values that could possibly range from 0.277~mHz to 100~mHz. In this case, the value of $\mathcal{W}$ from Eq.~(\ref{eq:IHSstat}) is diminished for higher values of $\Delta f_\text{obs}$. This results in upper limits that are dominated by those putative signals having large $\Delta f_\text{obs}$. One can show empirically that the upper limit values improve (smaller strain upper limits) as $(\Delta f_\text{obs})^{0.4}$ for smaller cutoffs in $\Delta f_\text{obs}$.

\subsection{\label{sec:outlierprocedure}All-sky outlier follow up results}
Outliers reported from the first stage of the pipeline, the IHS algorithm, are tested using the second stage in order to confirm or reject each candidate. Those outliers passing the second, template-based stage are ranked by their false alarm probability value, which indicates their significance of occurring in Gaussian noise alone. The false alarm probability of a weighted sum of $\chi^2$ variables is numerically determined using methods described in~\citep{Davies2,TwoSpectMethod}. Only outliers whose false alarm probability in a single detector is more significant than $10^{-18}$, corresponding to a Gaussian SNR of $\sim$8.8, are followed up as possible candidate gravitational wave signals. Such a strict threshold is set in order to reduce the number of outliers produced by non-Gaussian noise artifacts.

\begin{figure}
\centering
\includegraphics[width=0.5\textwidth]{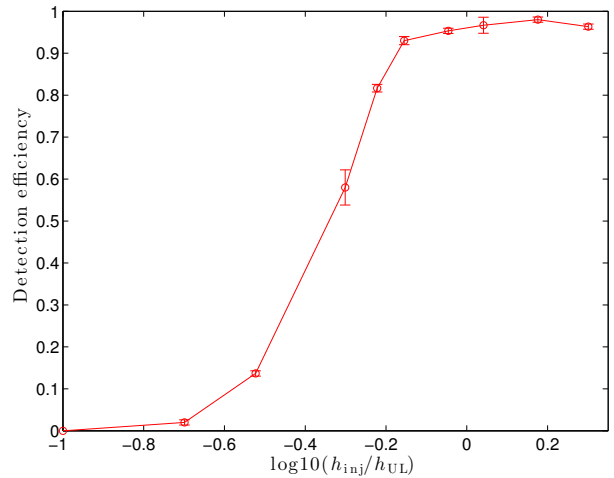}
\caption{\label{fig:coincidenceEfficiency}The efficiency of signals passing the given coincidence requirements as a function of the injection amplitude (normalized to the upper limit value at the specific frequency of the injection).}
\end{figure}

Those candidates from each detector with a fiducial signal frequency greater than 50 Hz are then subjected to multi-detector coincidence tests in the multi-dimensional search parameter space. Coincidence requirements were tested using simulated signals to determine the false dismissal probability as a function of the injected strain values (see figure~\ref{fig:coincidenceEfficiency}). The choice of coincidence requirements (see table~\ref{tab:coincidence}) are shown to be sensible given the false dismissal probability should on average be no greater than 5\% at the upper limit value (see figure~\ref{fig:coincidenceEfficiency}). Coincident candidates are required to have an orbital period difference, $dP$, that scales with period and modulation depth as
\begin{eqnarray}
dP <& (4.5T_{\rm obs})^{-1}\times\textrm{min}[P_1^2\left(\Delta f_1/3.6\,{\rm mHz}\right)^{-1/2},\nonumber \\*
 & P_2^2\left(\Delta f_2/3.6\,{\rm mHz}\right)^{-1/2}]\,,
\label{eq:Pmismatch}
\end{eqnarray}
where $P_1$ and $P_2$ are the two identified orbital period values and $\Delta f_1$ and $\Delta f_2$ are the two identified modulation depths of the outliers in each detector. The coincidence requirements also allow for harmonic confusion in $P_1$ or $P_2$ up to the third harmonic or sub-harmonic.
\begingroup
\squeezetable
\begin{table}
\caption{\label{tab:coincidence}Coincidence requirements for follow-up of outliers between two detectors.}
\begin{ruledtabular}
\begin{tabular}{lr}
Parameter & Allowed difference \\
\hline
Gravitational wave frequency mismatch & 0.556~mHz \\
Orbital period mismatch\footnote{Outliers may have improperly identified orbital period due to harmonic confusion. Three higher harmonics and three sub-harmonics are tested in addition to the fiducial orbital period.} & See Eq.~(\ref{eq:Pmismatch}) \\
Modulation depth mismatch & 0.556~mHz \\
Sky position mismatch & 0.1$(800\,{\rm Hz}/f)$~rad
\end{tabular}
\end{ruledtabular}
\end{table}
\endgroup

The observed loss in detection efficiency is the result of restricting the first stage of the pipeline to pass only the five most interesting outliers to the second stage of the pipeline in order to limit computational resources spent in the second stage. The limitation means that the five outliers can sometimes ($\sim$5\% of the time) have correlated offsets in their parameters from the true waveform parameters (for example, offset from the fiducial frequency and modulation depth by a correlated amount), but the second stage is unable to find the true parameters from this limited subset of outliers simultaneously in two or more detectors. On average, the false dismissal of a simulated, large amplitude signal is $\sim$5\%. Considering those simulations where a simulated signal with modulation depth $\Delta f$ is close to the maximum observable modulation depth using TwoSpect ($\Delta f/\Delta f_\text{max}\gtrsim0.3$, see figure~\ref{fig:paramspace}), the false dismissal probability increases above 5\%, approaching 50\% false dismissal at the highest values of $\Delta f$. On the other hand, the false dismissal probability falls below 5\% when the ratio $\Delta f/\Delta f_\text{max}$ becomes small ($\lesssim0.1$). The false dismissal probability of the outlier follow up analysis, however, does not affect the ability of the TwoSpect pipeline to set accurate upper limits on $h_0$.

Pair-wise combinations are made in each detector where for each outlier of the first detector only the most significant outlier of the second detector passing the coincidence requirements is retained. The same procedure is performed with the detector lists reversed. From this final list of outliers passing coincidence requirements, the outliers are grouped into narrow frequency bands (typically less than $\sim$30~mHz) for further, manual inspection. The pair-wise combination that has the smallest false alarm probability is considered representative of the outliers in each group~\footnote{The period difference term also allows for orbital frequency harmonic confusion up to the third harmonic or sub-harmonic in each detector.}. The sky position of the pair is then averaged and given in table~\ref{tab:outliers}.
\begingroup
\squeezetable
\begin{table*}
\caption{\label{tab:outliers}Most significant outliers passing coincidence requirements grouped by frequency.}
\begin{ruledtabular}
\begin{tabular}{D{.}{.}{3.6}D{.}{.}{3.6}D{.}{.}{4.6}D{.}{.}{4.6}D{.}{.}{3.3}D{.}{.}{3.3}D{.}{.}{1.4}D{.}{.}{1.4}ll}
\multicolumn{1}{c}{$f_1$} & \multicolumn{1}{c}{$f_2$} & \multicolumn{1}{c}{$P_1$} & \multicolumn{1}{c}{$P_2$} & \multicolumn{1}{c}{$\Delta f_{{\rm obs},\,1}$} & \multicolumn{1}{c}{$\Delta f_{{\rm obs},\,2}$} & \multicolumn{1}{c}{$\alpha$} & \multicolumn{1}{c}{$\delta$} &  &  \\
\multicolumn{1}{c}{{\rm (Hz)}} & \multicolumn{1}{c}{{\rm (Hz)}} & \multicolumn{1}{c}{{\rm (ks)}} & \multicolumn{1}{c}{{\rm (ks)}} & \multicolumn{1}{c}{{\rm (mHz)}} & \multicolumn{1}{c}{{\rm (mHz)}} & \multicolumn{1}{c}{{\rm (rad)}} & \multicolumn{1}{c}{{\rm (rad)}} & Pair & Cause \\
\hline
61.000000 & 60.999735 & 7896.759938 & 7896.662780 & 100.000 & 99.722 & 4.6694 & -1.5057 & H1, V1 & 61 Hz line \\
90.010000 & 90.010130 & 6303.365336 & 7383.886742 & 3.333 & 3.611 & 6.0104 & 0.7223 & H1, V1 & 90 Hz line \\
99.993333 & 99.993120 & 5406.840000 & 6337.503156 & 2.222 & 2.500 & 2.9800 & -1.0808 & H1, V1 & 100 Hz line, power line \\
108.856944 & 108.856944 & 86.166079 & 86.189136 & 0.278 & 0.556 & 3.1399 & -0.5979 & H1, L1 & Fake pulsar 3 \\
127.985321 & 127.985331 & 6190.885362 & 6180.988184 & 1.944 & 1.944 & 3.3945 & 0.0542 & H1, L1 & 128 Hz line \\
134.092913 & 134.093056 & 7241.303571 & 803.807379 & 0.833 & 0.556 & 4.1811 & 0.7177 & H1, L1 & 134.1 Hz line \\
192.498889 & 192.498889 & 8034.622934 & 7962.429248 & 94.444 & 93.889 & 2.3034 & 0.4624 & H1, L1 & Fake pulsar 8 \\
217.997855 & 217.998056 & 6418.038156 & 5406.840000 & 2.222 & 2.500 & 4.7035 & 1.2218 & H1, L1 & 218 Hz line \\
222.009771 & 222.010278 & 6208.893015 & 6337.503156 & 2.222 & 2.500 & 5.8797 & 1.2601 & H1, L1 & 222 Hz line \\
249.999167 & 249.999444 & 7704.013336 & 6268.095166 & 4.167 & 4.444 & 1.0903 & -1.3752 & H1, L1 & 250 Hz line \\
256.027734 & 256.027500 & 6181.988411 & 5406.840000 & 2.222 & 2.500 & 6.2214 & 0.0942 & H1, L1 & 256 Hz line \\
282.000000 & 282.000070 & 5497.564390 & 7526.367421 & 12.222 & 12.500 & 3.2944 & -1.4169 & H1, L1 & 282 Hz line \\
392.000139 & 392.000556 & 6424.556015 & 7546.515502 & 1.389 & 1.111 & 4.7405 & 1.2000 & H1, L1 & 392 Hz line \\
392.179468 & 392.179958 & 6179.525429 & 6144.240561 & 2.222 & 1.667 & 4.1648 & -1.4600 & H1, L1 & 392.2 Hz DAQ line \\
404.750625 & 404.750140 & 6187.506004 & 6187.506004 & 60.278 & 60.278 & 4.9828 & 1.4683 & H1, L1 & 404.7 Hz PCal line \\
410.000711 & 410.000278 & 7476.196898 & 7358.175495 & 24.722 & 24.167 & 0.3311 & 1.1811 & H1, L1 & 410 Hz line \\
413.000000 & 412.999583 & 7240.033559 & 7462.679595 & 80.000 & 80.000 & 0.8309 & -0.7947 & H1, L1 & 413 Hz line
\end{tabular}
\end{ruledtabular}
\end{table*}
\endgroup

All of the outliers listed in table~\ref{tab:outliers} are found to be associated with known detector artifacts~\citep{LIGODetchar,VirgoDetchar}. Most of the outliers are caused by a comb of 1~Hz, 2~Hz, or 16~Hz harmonics associated with the LIGO data acquisition system (DAQ). Another outlier is also due to a 392.2~Hz DAQ line. Two of the outliers are due to fake continuous gravitational wave signals with unrealistically large amplitudes injected into the detectors by modulating the interferometer arm lengths (see, e.g.,~\citep{FullS5EatH} for additional details). Another outlier is due to a photon calibrator~\citep{LIGOPcal} calibration line at 404.7~Hz observed in the gravitational wave data channel. One other outlier is caused by a narrow, previously unidentified spectral artifact in H1 at 134.1~Hz coinciding with noise fluctuations in L1 to produce a candidate signal. Further studies of this outlier have shown the signal characteristics are inconsistent with a gravitational wave signal. There are no TwoSpect outliers passing coincidence requirements in all of H1, L1, and V1 in the 50 to 100 Hz frequency band.

\subsection{Upper limits on Scorpius X-1 emission}
A separate, opportunistic analysis has been carried out for possible continuous gravitational wave emission from Sco X-1 using the same analysis as the all-sky search. In this second analysis, however, only the sky location of Sco X-1 was searched, and the parameter space was restricted to coincide with the projected semi-major axis $a\sin i=1.44\pm0.18$~ls~\citep{ScoX1asini} and $P=68023.70\pm0.0432$~s~\citep{ScoX1orbitP}. The highest frequency that can be searched given these parameters and assuming the gravitational wave signal is contained within a single SFT frequency bin, is given by~\citep{TwoSpectMethod}
\begin{eqnarray}
f &\leq& 78.9229\left(\frac{P}{68023.70\,\text{s}}\right)^2 \times \nonumber \\*
& &\left(\frac{a\sin i}{1.44\,\text{ls}}\right)^{-1}\left(\frac{T_\text{SFT}}{1800\,\text{s}}\right)^{-2}\,\text{Hz}\,.
\end{eqnarray}
Assuming a worst-case scenario of $a\sin i = 1.44+3\times0.18=1.98$~ls and $T_{\rm SFT}=1800$~s, this relation limits the highest frequency that can be searched to be 57.25~Hz because we analyze only full-0.25~Hz frequency bands. Note that there is good reason, however, to believe that the Scorpius X-1 signal frequency would be higher than this value~\citep{Wattsetal}. Using Eq.~(\ref{eq:moddepth2}), the range of $\Delta f_\text{obs}$ is frequency-dependent and ranges from 1.663~mHz to 10.470~mHz. Data from S6 were analyzed from 50~Hz up to 57.25~Hz, while VSR2/3 data were analyzed from 20~Hz up to 57.25~Hz.

\begin{figure}
\centering
\includegraphics[width=0.5\textwidth]{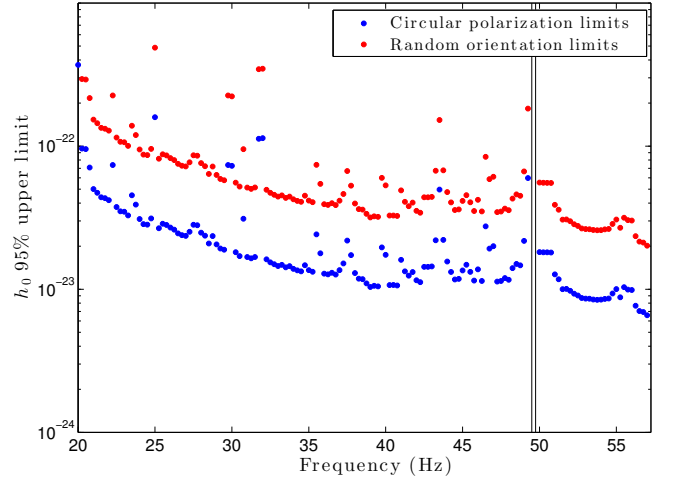}
\caption{\label{fig:scox1ulresults}Sco X-1 strain upper limit results of S6/VSR2-3 for continuous gravitational waves assuming the source waves are circularly polarized (blue points) or the source waves are randomly polarized with random pulsar orientations (red points). The black vertical lines indicate 0.25~Hz frequency bands in which no upper limits have been placed.}
\end{figure}

The combined upper limits of the three interferometers are shown in figure~\ref{fig:scox1ulresults}. The upper limit results are typically about a factor of 3 better than the all-sky upper limits in this frequency range because only a single sky location needs to be searched, the range of orbital parameters to be searched is much smaller, and the incoherent harmonic summing step used $S=10$ folds of the second FFT spectra as opposed to $S=5$ for the all-sky search. These results are comparable to results from the fifth LIGO Science Run~\citep{S5directionalStochastic} using a different analysis technique~\citep{S4directionalStochastic}.

A more sensitive search and more constraining upper limits could be obtained by optimizing the search pipeline for sources with known orbital parameters and unknown spin parameters. A future publication will detail these changes and demonstrate the improvement for such a search.

\begin{figure}
\centering
\includegraphics[width=0.5\textwidth]{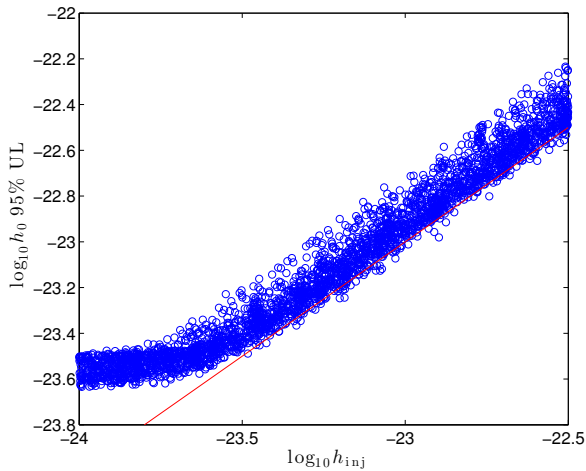}
\caption{\label{fig:ulvalidation}Results from an upper limit validation test at 401 Hz, for circularly-polarized waves. Each data point (blue circles) gives the 95\% confidence upper limit set by TwoSpect for a given amplitude of a circularly-polarized injection. The red line indicates a slope of 1.  The upper limit procedure is valid provided that no more than 5\% of the blue circle data points lie below the red line for any value of the injected amplitude.}
\end{figure}

\begin{figure}
\centering
\includegraphics[width=0.5\textwidth]{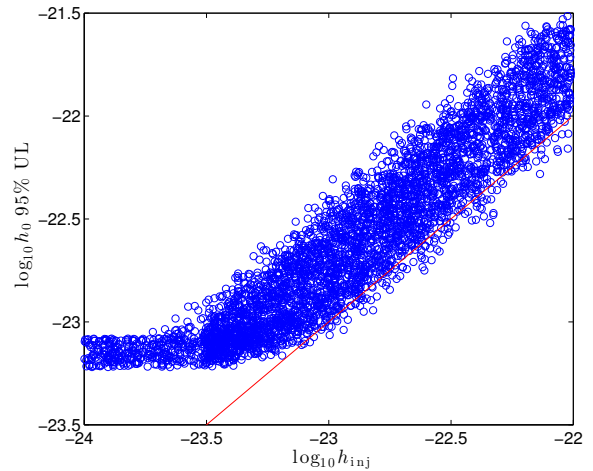}
\caption{\label{fig:randPolarizationUL}Results from an upper limit validation test at 401 Hz, for randomly-polarized waves. Each data point (blue circles) gives the 95\% confidence upper limit set by TwoSpect for a given amplitude of a randomly-polarized injection. The red line indicates a slope of 1.  The upper limit procedure is valid provided that no more than 5\% of the blue circle data points lie below the red line for any value of the injected amplitude.}
\end{figure}

The methodology used in obtaining both the all-sky and Sco X-1 upper limits on source strain has been validated with simulated signal injections. Figures~\ref{fig:ulvalidation} and \ref{fig:randPolarizationUL} illustrate validation tests for circularly and randomly polarized signals, and figure~\ref{fig:coincidenceEfficiency} shows derived detection efficiency.

\section{\label{sec:conclusions}Conclusions}
We have carried out the first explicit all-sky search for continuous gravitational wave signals from unknown spinning neutron stars in binary systems. This search was made possible through the use of the TwoSpect algorithm~\citep{TwoSpectMethod} to look for these sources. The search relies on the periodic modulation of the gravitational waves caused by the orbital motion of the source. The doubly Fourier-transformed data is processed by a hierarchical pipeline, subjecting the data to an incoherent harmonic summing stage, followed by comparing interesting regions of the data to templates that approximate the expected signal power in the doubly Fourier-transformed data.

This search has covered a broad range of possible gravitational wave frequencies---from 20~Hz to 520~Hz---binary orbital periods---from 2~h to 2,254.4~h---and frequency modulations---from 0.277~mHz to 100~mHz. These parameters cover a wide range of binary orbital parameters, and many known binary systems with neutron stars fall into the parameter space of this search. No plausible candidate continuous gravitational wave signal was observed. Upper limits are placed on continuous gravitational waves from unknown neutron stars in binary systems over the parameter spaced searched. The search carried out here is the most sensitive that covers such a wide range of the binary orbital parameter space~\citep{TwoSpectMethod}.

Additionally, we have carried out a search for continuous gravitational waves from Sco X-1 between 20~Hz and 57.25~Hz. This search has covered only a small range of possible spin frequencies of Sco X-1 because of the limitations from using 1800~s SFTs. To more fully cover the range of possible Sco X-1 spin frequencies, future searches will need to use shorter coherent length SFTs. No outliers passed the same thresholds used for the all-sky search. Upper limits are placed on gravitational wave emission from Sco X-1 using a dedicated pipeline that assumes a continuous wave model for the gravitational radiation emitted by the neutron star.

Second-generation gravitational wave detectors will have a broadband noise improvement by about a factor of 10. Using TwoSpect with data from second generation detectors will probe even deeper and wider regions of parameter space. Although originally developed for an all-sky search, the core TwoSpect pipeline could be tuned to be used as a directed search method for known binary systems with poorly constrained orbital parameters.

\begin{acknowledgements}
The authors gratefully acknowledge the support of the United States
National Science Foundation for the construction and operation of the
LIGO Laboratory, the Science and Technology Facilities Council of the
United Kingdom, the Max-Planck-Society, and the State of
Niedersachsen/Germany for support of the construction and operation of
the GEO600 detector, and the Italian Istituto Nazionale di Fisica
Nucleare and the French Centre National de la Recherche Scientifique
for the construction and operation of the Virgo detector. The authors
also gratefully acknowledge the support of the research by these
agencies and by the Australian Research Council, 
the International Science Linkages program of the Commonwealth of Australia,
the Council of Scientific and Industrial Research of India, 
the Istituto Nazionale di Fisica Nucleare of Italy, 
the Spanish Ministerio de Econom\'ia y Competitividad,
the Conselleria d'Economia Hisenda i Innovaci\'o of the
Govern de les Illes Balears, the Foundation for Fundamental Research
on Matter supported by the Netherlands Organisation for Scientific Research, 
the Polish Ministry of Science and Higher Education, the FOCUS
Programme of Foundation for Polish Science,
the Royal Society, the Scottish Funding Council, the
Scottish Universities Physics Alliance, The National Aeronautics and
Space Administration, 
OTKA of Hungary,
the Lyon Institute of Origins (LIO),
the National Research Foundation of Korea,
Industry Canada and the Province of Ontario through the Ministry of Economic Development and Innovation, 
the National Science and Engineering Research Council Canada,
the Carnegie Trust, the Leverhulme Trust, the
David and Lucile Packard Foundation, the Research Corporation, and
the Alfred P. Sloan Foundation. This article has LIGO document number LIGO-P1300048.
\end{acknowledgements}

\bibliographystyle{apsrev4-1}
\bibliography{references}

\end{document}